# High-Conductance, Ohmic-like HfZrO$_4$ Ferroelectric Memristor

Laura Bégon-Lours[†,1], Mattia Halter[1,2], Youri Popoff[1,2], Zhenming Yu[1,3], Donato Francesco Falcone[1,4], Bert Jan Offrein[1]

[1]Neuromorphic Devices and Systems, IBM Research, Rüschlikon, Switzerland
[2]ETH Zürich, Zürich, Switzerland,
[3]Institute of Neuroinformatics, University of Zürich & ETH Zürich, Zürich, Switzerland,
[4]EPFL, Lausanne, Switzerland
{[†]lbe ; att ; ypo ; zyu ; dfa ; ofb}@zurich.ibm.com

*Abstract*— The persistent and switchable polarization of ferroelectric materials based on HfO$_2$-based ferroelectric compounds, compatible with large-scale integration, are attractive synaptic elements for neuromorphic computing. To achieve a record current density of 0.01 A/cm$^2$ (at a read voltage of 80 mV) as well as ideal memristive behavior (linear current-voltage relation and analog resistive switching), devices based on an ultra-thin (2.7 nm thick), polycrystalline HfZrO$_4$ ferroelectric layer are fabricated by Atomic Layer Deposition. The use of a semiconducting oxide interlayer (WO$_{x<3}$) at one of the interfaces, induces an asymmetric energy profile upon ferroelectric polarization reversal and thus the long-term potentiation / depression (conductance increase / decrease) of interest. Moreover, it favors the stable retention of both the low and the high resistive states. Thanks to the low operating voltage (<3.5 V), programming requires less than 10$^{-12}$ J for 20 ns long pulses. Remarkably, the memristors show no wake-up or fatigue effect.

## I. Introduction

The use of HfO$_2$-based ferroelectric compounds [1], in particular HfZrO$_4$ (HZO) [2], receive increasing interest for the fabrication of hardware for neuromorphic and in-memory computing. For example, HfO$_2$ based ferroelectric field-effect transistors can implement logic or static random access memory functionalities at the nanoscale [3], [4] [5]. Moreover, the relatively low crystallization temperature required to obtain the ferroelectric phase enables back-end-of-line compatibility, e.g. for the fabrication of three-terminal tunable resistances for synaptic weights [6]. Such devices, as well as two-terminal memristors [7] are the building blocks of artificial neural network accelerators [8], [9], where the matrix-vector multiplication is implemented in the analog domain by a parallel accumulation of currents generated across an array of tunable resistances. Ferroelectric memristors are based on a ferroelectric layer separating two electrodes of a different material, thin enough to allow electrical conduction and thick enough to stabilize ferroelectricity. The energy profile of the device is modified upon polarization reversal, inducing the long-term modification of the resistance. This effect is demonstrated with ferroelectric perovskites[10]–[13] as well as recently with epitaxial HfZrO$_4$ (HZO) [14], [15]. These conditions have also been obtained for polycrystalline HZO [16]–[22], and under a significant bias it is possible to measure through HZO barriers as thick as 10 nm [23]–[25]. However, due to the insulating nature of HZO, these devices have inherently small current densities and their footprint is thus relatively large. State-of-the-art double-layer HfO$_2$(10 nm)/Al$_2$O$_3$ junctions [26], [27] are typically measured at a bias of 2 V, for which the ratio between the current in the On and Off state (On/Off ratio) is as high as 10. The disadvantages of these junctions are the low current density (~ 1uA/cm$^2$), a poor retention of the high resistive state due to charge trapping in the Al$_2$O$_3$ layer, and the non-linear current-voltage characteristic, which makes the use of a logarithmic driver compulsory (as in [28]) for analog vector-matrix multiplication operations. In this work, we exploit a different junction concept: an ultra-thin HZO layer with a WO$_x$ interlayer at one of the electrodes. The choice of WO$_x$ is motivated by two aspects. First, it is an oxide, as HZO, ensuring that polarization charge screening occurs at the HZO/WO$_x$ interface. This avoids trapping / detrapping of charges at an uncontrolled interfacial oxide or in a dielectric layer, which would deteriorate the endurance and retention properties. Second, substochiometric WO$_x$, x<3, is a n-type semiconductor, with carrier densities in the range of 10$^{16}$~10$^{21}$ cm$^{-3}$[29]. At the HZO/WO$_x$ interface, an accumulation (resp. depletion) of carriers in the WO$_x$ occurs when the polarization points towards (resp. outwards) the WO$_x$ layer, ensuring an asymmetric energy profile for the junction. At the cost of a reduced On/Off ratio, the 2.7 nm thick HZO layer enables the devices to operate at low bias (<100 mV), in a linear regime, with a large current density of 0.01 A/cm$^2$ at 80 mV (1 A/cm$^2$ at 1V). Such device operating in the quasi-ohmic regime was already demonstrated in HZO/TiO$_x$ junctions [16], where the On/Off ratio was 2 and the current density was only 1 uA/cm$^2$ at a read voltage of 80 mV. This work further establishes the long-term depression / potentiation capability of ferroelectric, two-terminals synaptic weights.

## II. Fabrication

The fabrication process uses techniques compatible with large-scale, back-end-of-line integration. A stack comprising of a 20 nm thick bottom electrode of TiN, a 2 nm thick interlayer of WO$_x$, a 2.7 nm thick HZO and a 10 nm thick TiN top electrode is grown by Atomic Layer

Deposition (ALD) on a conducting Si substrate. The ALD process does not exceed 375°C, and X-Ray characterization shows that the layers are amorphous after deposition. The native $SiO_2$ oxide at the surface of the wafer was removed by a dip in buffered hydrofluoric acid prior to the ALD growth. The crystallization is performed using the ms-Flash Lamp Annealing technique [30]: the film is pre-heated to a temperature of 500°C, then a flash of 90 J/cm$^2$ is applied during 20 ms. The HZO thickness of 2.7 nm is determined by X-Ray Reflectivity. Devices are fabricated by subtractive manufacturing using optical lithography. First a 100 nm thick tungsten layer is sputtered, then circular capacitors of diameters from 20 to 140 um are structured using Reactive Ion Etching: after etching the W layer, a thirty seconds Argon etch step removes the native oxide at the surface of the top TiN layer, which is then etch down to the HZO layer.

### III. LONG-TERM POTENTIATION / DEPRESSION

The synaptic functionality of the devices is characterized using a B1530A Waveform Generator / Fast Measurement Unit (WGFMU) and a SMU (Source Measurement Unit). The shared bottom electrode ($WO_x$/TiN/Si++) is grounded. In this section, we address both the programmability of the device for inference and its response under pulses of varying sign, amplitude and width, emulating the combination of pre-synaptic pulses (negative bias) and post-synaptic pulses (positive bias), described for example in [31] or [32].

First, the "weight" or conductance of the synapse is modulated by pulses of constant duration (20 ns) and increasing or decreasing amplitude. After each pulse of amplitude $V_{write}$ (in abscise of Figure 1 a)) a DC current-voltage (I-V) measurement is performed to get the resistance at $V_{read}$ = 0.1 V. Upon positive bias from $V_{write}$ = 0 to 0.8 V, the resistance of the device is constant. Then, for pulse amplitudes of 0.8 to 2.8 V, it increases (synapse long-term depression). Operating with 20 ns long pulses allows to program with an energy in the pico-Joule range. Thanks to the small reading voltage of 0.1 V, the reading operation dissipates less than a femto-Joule for an integration time below a micro-second. Pulses of the same sign but with decreasing amplitude allows several micro-amperes to flow through the junction but only slightly modify the resistance (hysteresis, from 2.8 V to -0.8 V), showing that the resistive switching is field- and not current-driven, consistent with a mechanism originating from the ferroelectric properties of the HZO layer. The small drift of the resistance upon reading is due to the voltage level that is very close to the coercive voltage of the device in DC mode. Reciprocally, upon pulses of negative sign and amplitudes larger than 0.4 V, the resistance of the artificial synapse decreases. Cycles are reproducible and intermediate states are stable, as shown by the minor loops in Figure 1 a). Remarkably, the I-V characteristics of the junction (see Figure 1 b)) are linear around the read voltage, which makes these devices ideal memristors and excellent devices for analog multiplication.

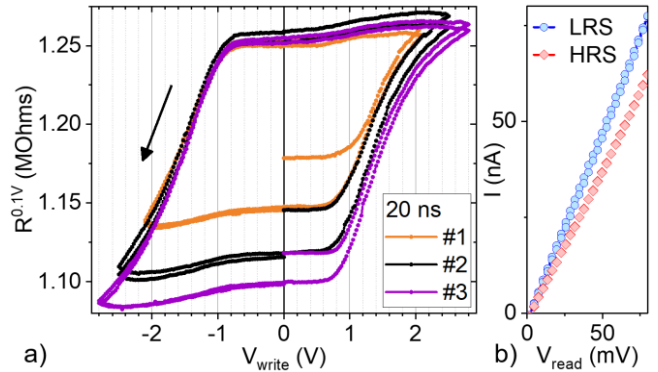

**Figure 1:** a) Resistance of the device, measured at 0.1 V, after pulses of constant duration (20 ns) and increasing amplitude $V_{write}$. The arrow indicates the chronological occurrence. b) High-resolution current-voltage (I-V) read after applying 1 $V_{DC}$ (High Resistive State, HRS) and -1 $V_{DC}$ (Low Resistive State, LRS) showing linear behavior.

Long-term potentiation / depression of the device can also be achieved with longer pulses. As the duration of the pulse increases, the voltage required to operate the dynamic range of the device decreases: Figure 2 a) shows the resistance measured at $V_{read}$ = 0.1 V after pulses of constant duration $t_{write}$ (see legend) and increasing amplitude $V_{write}$.

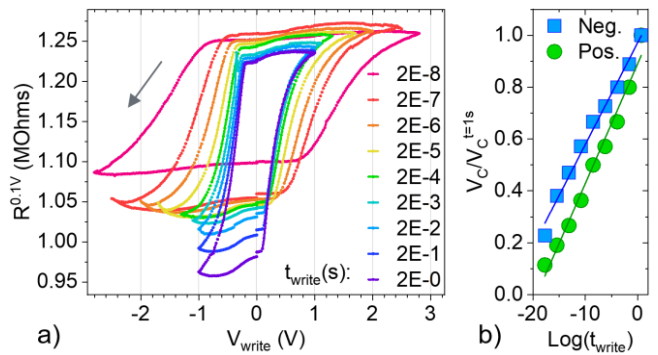

**Figure 2:** a) Resistance of a device, measured at 0.1 V, after pulses of constant duration $t_{write}$ (see legend, 20 ns to 2 s) and increasing amplitude $V_{write}$. The arrow indicates the chronological occurrence. b) Normalized coercive voltage for the positive bias (green circles) and negative bias (blue squares) and the corresponding linear regressions.

A straightforward explanation could be the capacitive behavior of both the $WO_x$ and the HZO layer. However, a linear relation exists between the coercive field $V_c$ and the logarithm of the pulse duration $t_{write}$ (see Figure 2 b)), following the empirical Merz's law describing ferroelectric switching dynamics [33]: $t_s \sim t_0 e^{-E_a/E}$ with $t_s$ the mean switching time, $E_a$ the activation field and $t_0$ a time parameter. This relation is a second indication that the resistive switching originates from ferroelectric effects.

Long-term potentiation and depression are also obtained by schemes with a constant amplitude and an increasing duration. Figure 3 shows the resistance of the device measured at 0.1 V (blue data points) after each pulse of duration $t_{write}$ (upper panel). The resistance increases (depression) when the amplitude of the pulses is 1 V and decreases (potentiation) when the amplitude of the pulses is -1 V. 20% of the dynamic range is covered by the first pulse (20 ns) consistently with the fast dynamics of ferroelectric switching. As for the scheme of constant pulse width and increasing duration, the long-term potentiation / depression is analog, reproducible from cycle to cycle, and symmetric.

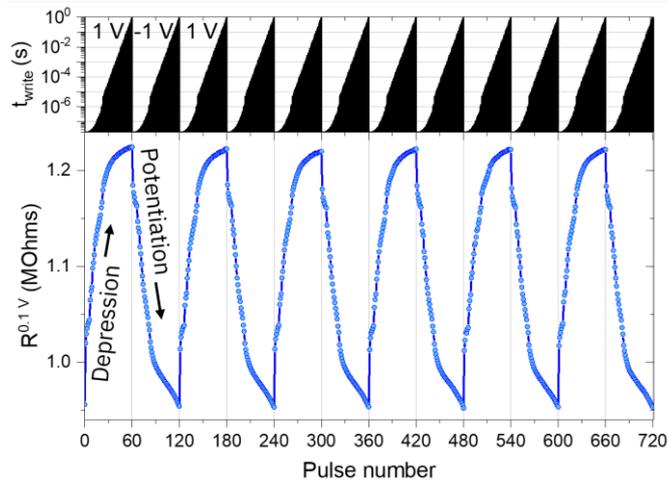

**Figure 3:** Resistance of a device measured at 0.1 V (blue data points) after each pulse of duration $t_{write}$ (upper panel). The amplitude of the pulse is 1 V for pulses (0-59), -1 V for pulses (60-119), and the pattern is reproduced each 120 pulses.

## IV. RETENTION AND ENDURANCE

The retention of the device under strong stress (DC bias) is measured for two weeks (see Figure 4). Three devices are swept with a DC bias from -1 V to 1 V and to -1 V again (Set / Reset sequence). Then a device is programmed in the intermediate state by a DC bias of 0.2 V, and another device in the HRS by a DC bias of 1 V. The resistance is measured at 80 mV during the first hour, and then one, two, and three weeks after.

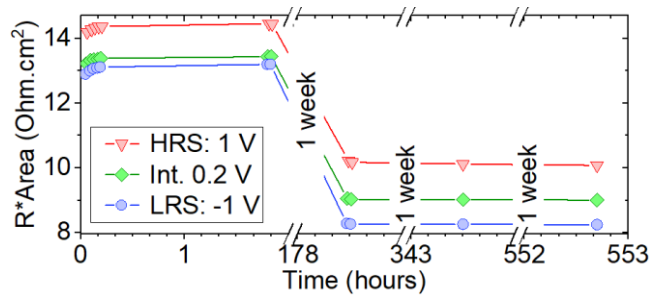

**Figure 4:** Retention of three devices programmed by a DC set/reset sequence (-1V to 1V to -1V) followed by a DC sweep and from 0 to 1 V (red triangles, HRS), or from 0 to 0.2 V (green diamonds, intermediate state), or not followed by a sweep (blue circles, LRS).

During the first minutes a relaxation occurs: the three devices see their resistance increase by the same magnitude. The resistance is then stable for the first hour. After one week, the three devices have drifted towards a lower resistance. The resistances measured after two and three weeks are identical to the resistances measured after one week, remarkably both for the intermediate and the high resistance state, showing that the depolarization field is not destabilizing the device despite the reduced thickness of the ferroelectric layer [34]. This retention at long time scales makes these devices good candidate for inference.

Furthermore, the memristors show only little wake-up or fatigue effect, which especially facilitates online learning applications. Figure 5 a) shows the I-V characteristics of devices that have been cycled at +/- 1V, at a frequency of 100 kHz, $10^0$ (pink data) up to $10^8$ times (purple data), set in the high resistive states (dark data points) and then in the low resistive state (light data points). Upon fatigue, there is a slight decrease of the resistance but no deterioration of the On/Off ratio. Similarly, Figure 5 b) shows the full I-V sweeps for the same devices, also showing minor modification upon fatigue.

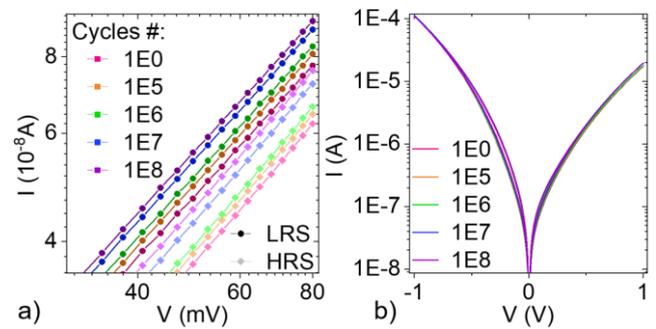

**Figure 5**: I-V characteristics of devices having been cycled at +/- 1V, at a frequency of 100 kHz, $10^0$ (pink data) up to $10^8$ times (purple data) a) after being set in the high resistive states (dark data points) and then in the low resistive state (light data points); b) full sweep.

## V. CONCLUSION

The memristive properties of a ferroelectric two-terminals devices are studied in the context of in-memory computing for neuromorphic hardware. Programming, as well as long-term potentiation / depression is achieved by modulating the sign, the amplitude and the duration of the pulses, emulating pre- and post-synaptic activity. Compared to state-of-the-art junctions, the proposed devices show improved scalability and operate in the Ohmic-like regime, ideal for analog multiplication but suffer from a reduced On/Off ratio. The devices show good retention properties and stable behavior upon fatigue.

ACKNOWLEDGMENT

This work is funded by H2020: FREEMIND (n° 840903), ULPEC (n°732642) and BeFerroSynaptic (n° 871737). The authors acknowledge the *Binnig and Rohrer Nanotechnology Center* (*BRNC*).